\documentstyle[12pt]{article}
\begin{document}
\def\beq{\begin{equation}}
\def\eeq{\end{equation}}
\def\bea{\begin{eqnarray}}
\def\eea{\end{eqnarray}}
\def\ve{\vert}
\def\vel{\left|}
\def\ver{\right|}
\def\nnb{\nonumber}
\def\ga{\left(}
\def\dr{\right)}
\def\aga{\left\{}
\def\adr{\right\}}
\def\rar{\rightarrow}
\def\nnb{\nonumber}
\def\la{\langle}
\def\ra{\rangle}
\def\ba{\begin{array}}
\def\ea{\end{array}}
\def\tep{$B \rar K \ell^+ \ell^-$}
\def\tepm{$B \rar K \mu^+ \mu^-$}
\def\tept{$B \rar K \tau^+ \tau^-$}
\def\ds{\displaystyle}

\title{ {\small {\bf 
EXCLUSIVE RADIATIVE WEAK DECAYS OF $B_c$ MESON IN LIGHT CONE QCD } } }

\author{ {\small T. M. AL\.{I}EV$^{a,}$ \footnote{taliev@cc.emu.edu.tr}\,,
M. SAVCI$^{b,}$ \footnote{savci@newton.physics.metu.edu.tr} }\\
{\small a) Physics Department, Girne American University} \\
{\small Mersin--10, Turkey}\\
{\small b) Physics Department, Middle East Technical University} \\
{\small 06531 Ankara, Turkey} }

\date{}

\begin{titlepage}
\maketitle
\thispagestyle{empty}

\begin{abstract}
\baselineskip  0.7cm
We investigate the radiative $B_c \rar \rho^+\gamma$ and $B_c \rar
K^{*+}\gamma$ decays in the standard model. The transition form
factors are calculated in the framework of the light cone QCD sum 
rules method. We estimate the branching ratios of the 
$B_c \rar \rho^+\gamma$ and $B_c \rar K^{*+}\gamma$ decays.

\end{abstract}

\vspace{1cm}
\end{titlepage}

\section{Introduction}

The experimental and theoretical investigation of the heavy flavored
hadrons is one of the most promising research area in high energy physics. 
These investigations might shed light for a precise determination of the
many fundamental parameters of the standard model (SM), such as
Cabibbo--Kobayashi--Maskawa (CKM) matrix elements, leptonic decay constants
of the heavy mesons 
and for deeper understanding of the dynamics of QCD. From this point
of view $B_c$ mesons are very interesting particles since in
their decay processes, different mechanisms (weak annihilation, charged
current, spectator and
FCNC decays) can give contributions simultaneously and this circumstance 
can play central role for understanding 
the dynamics of weak decays of heavy hadrons. 
Moreover the decay channels of the $B_c$
mesons are richer than $B$ meson decays and, since they contain two heavy
quarks, their QCD predictions are much more reliable. The study of the $B_c$
mesons, for these reasons, receives special attention among researches.

Note that,
the possibility of the production of the $B_c$ mesons in different colliders
and their different decay channels have already been extensively discussed
in the current literature \cite{R1}.
Among all different decay channels, the weak exclusive
radiative decays of the $B_c$ mesons play potentially very important role 
for the determination of the CKM parameters, similar to
the radiative $B$ meson decays.
In the usual $B_{u,d,s} \rar V \gamma~(V=\rho^+,~K^{*+})$ 
decays, two different mechanisms, namely weak annihilation and FCNC,
contribute simultaneously. Therefore extracting information about the CKM
matrix elements would involve a trustworthy estimate of both contributions
to the decay amplitude. Since, in calculation of the $B \rar V$ matrix
element, both contributions involve uncertainties of
their own, the resulting error in our attempt to estimate the CKM
parameters may be substantial.   

In contrast to the above--mentioned $B_{u,d,s} \rar V \gamma$ decay, the 
$B_c \rar V \gamma$ process is described only through the weak annihilation
mechanism. Therefore, investigation of the exclusive radiative 
weak $B_c$ meson decays is more reliable and promising 
in determination of the CKM parameters.

In this work we study the $B_c \rar V \gamma ~\ga V=\rho^\pm,~K^{*\pm} \dr$
decay in the SM, in the framework of the QCD sum rules. The paper is
organized as follows. In Section 2 we calculate the transition form factors
for the $B_c \rar V \gamma$ decay decay in the light cone QCD sum rules
approach. Section 3 is devoted to the numerical analysis and discussion of
our results. 

\section{Sum rules for transition form factors}

The relevant effective Hamiltonian for the $B_c \rar V \gamma$
process is 
\bea
{\cal H} = \frac{G}{\sqrt{2}}\, a_1 V_{cb} V_{uq}^* \, \bar q \gamma_\mu
\ga 1 - \gamma_5 \dr u \, \bar c \gamma_\mu \ga 1 - \gamma_5 \dr b~,
\eea
where $q=d$ or $s$ and $V_{uq}$ represent the corresponding matrix
elements, i.e., $V_{ud}$ or $V_{us}$, and the factor $a_1$ takes into
account renormalization of four fermion operators and it is numerically equal
to 1.13. In further analysis we will take $a_1 = 1$.
The matrix element for the above mentioned decay is 
\bea
{\cal M} = \frac{G}{\sqrt{2}}\,V_{cb} V_{uq}^* \,
\la V \gamma \vel \bar q \gamma_\mu \ga 1 - \gamma_5 \dr u \,
\bar c \gamma_\mu \ga 1 - \gamma_5 \dr b \ver B_c \ra ~.
\eea
In the factorization approximation, one may write this matrix element as
\bea
\lefteqn{
\la V \gamma \vel \bar q \gamma_\mu \ga 1 - \gamma_5 \dr  u \,
\bar c \gamma_\mu \ga 1 - \gamma_5 \dr b \ver B_c \ra = } \nnb \\
&&\la V \vel \bar q \gamma_\mu \ga 1 - \gamma_5 \dr u \ver 0 \ra
\la \gamma \vel \bar c \gamma_\mu \ga 1 - \gamma_5 \dr b \ver B_c \ra +
\la V \gamma \vel \bar q \gamma_\mu \ga 1 - \gamma_5 \dr u \ver 0 \ra
\la 0 \vel \bar c \gamma_\mu \ga 1 - \gamma_5 \dr b \ver B_c \ra~. 
\nnb \\ \nnb \\ 
\eea
Using the definitions
\bea
\la 0 \vel \bar c \gamma_\mu \gamma_5 b \ver B_c \ra &=&
-i \, f_{B_c} \ga p_{B_c} \dr_\mu \nnb \\
\la V \vel \bar q  \gamma_\mu \ga 1 - \gamma_5  \dr u  \ver 0  \ra &=&
\varepsilon_\mu^{(V)} m_V f_V~,
\eea
where $\varepsilon^{(V)}$, $f_V$ and $m_V$ are the polarization vector,
leptonic decay constant and mass of the vector $V$ meson, respectively,
one can easily show that the second term on the right side of Eq. (3) is
proportional to the light quark mass $m_q$, whose contribution is very
small (for more detail see \cite{R4} and \cite{R5}), 
and therefore we shall neglect it in
further analysis. Thus we conclude that the main contribution 
to $B_c \rar  V \gamma$
decay comes from the diagrams where photon is emitted from initial $b$ 
and $c$ quark lines. The corresponding
matrix element for the $B_c \rar  V \gamma$ decay can be written as 
\bea
{\cal M} = \frac{G}{\sqrt{2}}\, V_{cb} V_{uq}^* \varepsilon_\mu^{(V)}
m_V f_V \la \gamma \vel \bar c \gamma_\mu \ga 1 - \gamma_5 \dr b \ver
B_c \ra~.
\eea
All needs to be done then, is to calculate the matrix element
$\la \gamma \vel \bar c \gamma_\mu \ga 1 - \gamma_5 \dr b \ver B_c \ra$,
which describes the annihilation of the $B_c$ meson into 
$\bar c \gamma_\mu \ga 1 - \gamma_5 \dr b$ current with emission of a 
real photon. This matrix
element can be written in terms of the two independent, gauge invariant
(with respect to the electromagnetic interaction) structure as
\bea
\lefteqn{
\la \gamma (q) \vel \bar c \gamma_\mu \ga 1 - \gamma_5 \dr b \ver 
B_c(p+q) \ra = } \nnb \\
&& \sqrt{4 \pi \alpha} \Bigg\{\epsilon_{\mu\alpha\beta\sigma}    
e^{*\alpha} p^\beta q^\sigma \frac{g_1(p^2)}{m_{B_c}^2} +
i \, \left[e_\mu^* \ga p q \dr - \ga e^* p \dr q_\mu \right]
\frac{g_2(p^2)}{m_{B_c}^2} \Bigg\}~,
\eea
where $e^*$ and $q$ are the polarization vector and momentum of
the photon, respectively, $p$ is the momentum transfer 
$\ga p^2=m_V^2 \dr$, $g_1(p^2)$ and $g_2(p^2)$ are the parity conserving 
and parity
violating form factors. At this point we consider the problem of evaluating
the above--mentioned form factors, for which we will employ the light cone 
QCD sum rules approach (see the recent review
\cite{R3}). For this purpose we start by considering the following
correlator function
\bea
\Pi_\mu (p,q) = i \, \int d^4 x e^{ipx} \left< \gamma (q) \vel
{\cal T} \left\{ \bar c (x) \gamma_\mu \ga 1- \gamma_5 \dr b(x) \,
\bar b (0) i \gamma_5 c(0) \right\} \ver 0 \right>~.
\eea
This function can be decomposed into two independent structures, Lorentz and
gauge invariant, as follows:
\bea
\Pi_\mu (p,q) = \Pi_1 \epsilon_{\mu\alpha\beta\sigma} 
e^{*\alpha} p^\beta q^\sigma + i \, 
\Pi_2 \left[ e^*_\mu(p q)- (e^* p ) q_\mu \right]~.
\eea
In deep--Euclidean region $\ga p+q \dr^2 <0$ and $p^2=m_V^2
<\!\!<m_Q^2$, the heavy quarks $Q$ are far off--shell. Therefore photon emission
from the heavy quarks takes place perturbatively. This behavior of the 
$B_c \rar V \gamma$ decay is essentially different from the corresponding 
$B^\pm \rar V^\pm \gamma$ channel. In the latter the photon interacts
with quarks both perturbatively and non--perturbatively (see for example
\cite{R4,R5}), while in the $B_c \rar V \gamma$ decay the photon interacts
with quarks only perturbatively. 

Firstly let us calculate the physical part of the correlator (8). Inserting
the hadronic states with the relevant $B_c$ meson quantum numbers into 
Eq. (8) we get
\bea
\Pi^{(1,2)} = \frac{f_{B_c} m_{B_c}^2}{m_b+m_c} \, 
\frac{g_{1,2}}{\left[m_{B_c}^2 - (p+q)^2\right]} +
\int_{s_0}^\infty ds \,\frac{\rho^{(1,2)}(s,p^2)}
{s - (p+q)^2}~,
\eea
where we have used 
\bea
\left< B_c \vel \bar b \, i \gamma_5  c \ver 0 \right> =
\frac{f_{B_c} m_{B_c}^2}{m_b+m_c}~, \nnb
\eea
in Eq. (7). The second term in Eq. (9) represents contribution of the higher 
states starting from some threshold $s_0$. We invoke the hadron quark
duality and replace the hadron spectral density $\rho^h$ by the imaginary
part of the dispersion relation $\Pi^{(1,2)}$ calculated in QCD.

From (9) it follows that $\Pi^{(1,2)}\ga (p+q)^2,p^2 \dr$ is analytic in the
cut $(p+q)^2$ plane. In other words, in order to relate $\Pi^{(1,2)}$ with
its imaginary part we need dispersion relation in the variable $(p+q)^2$.
Therefore the perturbative contribution to the parity conserving and parity
violating amplitudes can be calculated by writing the dispersion integral in
the variable $(p+q)^2$, i.e., 
\bea
\Pi^{(1,2)} = \int ds \,\frac{\rho^{(1,2)}(s,p^2)}
{s - (p+q)^2} + \mbox{\rm subst. terms}~,
\eea
where the superscript 1 and 2 corresponds to $\Pi^{(1)}$ and $\Pi^{(2)}$
respectively, and $\rho^{(1,2)}$ are the spectral densities. These spectral
densities are calculated by a method presented in \cite{R6} (for
applications of this method, see also \cite{R7} and \cite{R8}). The
above--mentioned spectral densities were calculated in \cite{R9} in regard
to an investigation of the $B_c \rar \ell \nu \gamma$ decay and we shall
make use of these results, which lead to the following expressions for
$\Pi^{(1,2)}$:
\bea
\lefteqn{
\Pi^{(1)} = \sqrt{4 \pi \alpha} \, 
\frac{N_c}{4 \pi^2} \int \frac{ds}{\left[s-\ga p+q \dr^2\right]
\left[s-p^2\right]} } \nnb \\
&&\times
\Bigg\{ \ga m_b -m_c \dr \lambda \ga Q_c -Q_b \dr + Q_b m_b \, \mbox{ln}
\frac{1+\alpha-\beta+\lambda}{1+\alpha-\beta-\lambda}
+Q_c m_c \, \mbox{ln}
\frac{1-\alpha+\beta+\lambda}{1-\alpha+\beta-\lambda} \Bigg\}~, 
\\ \nnb \\ \nnb \\
\lefteqn{
\Pi^{(2)} = \sqrt{4 \pi \alpha} \, 
\frac{N_c}{4 \pi^2} \int \frac{ds}{\left[ s-\ga p+q \dr^2 \right]
\left[ s-p^2 \right]^2}} \nnb \\
&&\times \Bigg\{
m_b  Q_b \left[ \ga 2 m_b^2 + p^2 - s \dr \mbox{ln}
\frac{1+\alpha-\beta+\lambda}{1+\alpha-\beta-\lambda} -
\lambda \ga 2 m_b^2 - 2 m_c^2 + p^2 (2 -\alpha +\beta) -s \dr \right] 
\nnb \\
&&+ m_c  Q_b \left[ - 2 m_b^2 \, \mbox{ln}
\frac{1+\alpha-\beta+\lambda}{1+\alpha-\beta-\lambda}+
\lambda \ga 2 m_b^2 - 2 m_c^2 - p^2 (\alpha -\beta) + s \dr \right] \nnb \\
&&+ m_b  Q_c \left[ 2 m_c^2 \, \mbox{ln}
\frac{1-\alpha+\beta+\lambda}{1-\alpha+\beta-\lambda}+
\lambda \ga 2 m_b^2 - 2 m_c^2 - p^2 (\alpha -\beta) - s \dr \right]\\
&&+ m_c  Q_c \left[ \ga s - p^2 - 2 m_c^2 \dr \mbox{ln}
\frac{1-\alpha+\beta+\lambda}{1-\alpha+\beta-\lambda}- 
\lambda \ga 2 m_b^2 - 2 m_c^2 - p^2 (2 + \alpha -\beta) + s \dr \right]
\Bigg\} ~, \nnb
\eea
where $N_c=3$ is the color factor, $\alpha=m_b^2/s$, $\beta=m_c^2/s$. $Q_b$
and $Q_c$ are
the electric charges of the $b$ and $c$ quarks, respectively and
$ \lambda = \sqrt{1 +\alpha^2 + \beta^2 - 2 \alpha - 2 \beta
-2 \alpha \beta}$. 
Note that, as a formal check, when we set the charm quark mass $m_c$ to zero
in Eqs. (11) and (12), the resulting expressions are expected to be the
same as the ones calculated
for the perturbative part of the $B^\pm \rar V \gamma$ decay. 
This decay was investigated in
\cite{R4,R5}, and indeed our results for $\Pi^{(1,2)}$ 
coincide with theirs in the $m_c \rar 0$ limit.

The light cone QCD sum rule is obtained, as usual, by
equating the hadronic representation of the correlator $\Pi_\mu$ 
(see Eq. (9)) to the results  obtained through QCD calculations
(see Eqs. (11) and (12)). Applying Borel transformation in the variable 
$(p+q)^2$ to suppress the higher states, we get sum rules for the transition
form factors $g_1$ and $g_2$:
\bea
\lefteqn{
g_1(p^2) = \frac{m_b+m_c}{f_{B_c}} \, \frac{N_c}{4 \pi^2} 
\int_\Delta^1 \frac{du}{u} \,
e^{\left[m_{B_c}^2 u-\ga m_b+m_c \dr^2 + p^2 \bar u \right]/ \ga M^2 u \dr} }
\nnb \\
&&\times \Bigg[ \ga m_b - m_c \dr \lambda \ga Q_c - Q_b \dr + 
Q_b m_b \, \mbox{ln} \frac{1+\alpha-\beta+\lambda}{1+\alpha-\beta-\lambda} +
Q_c m_c \, \mbox{ln} \frac{1-\alpha+\beta+\lambda}{1-\alpha+\beta-\lambda}  
\Bigg]~,~~~~~ \\ \nnb \\ \nnb \\ \nnb \\
\lefteqn{
g_2(p^2) = \frac{m_b+m_c}{f_{B_c}} \, \frac{N_c}{4 \pi^2} \int_\Delta^1
\frac{du}{\left[ \ga m_b + m_c \dr^2 - p^2 \right]} \,
e^{\left[m_{B_c}^2 u-\ga m_b+m_c \dr^2 + p^2 \bar u \right]/ \ga M^2 u \dr} } 
\nnb \\
&&\times \Bigg\{
m_b  Q_b \left[ \ga 2 m_b^2 + p^2 - s \dr \mbox{ln}
\frac{1+\alpha-\beta+\lambda}{1+\alpha-\beta-\lambda} -
\lambda \ga 2 m_b^2 - 2 m_c^2 + p^2 (2 -\alpha +\beta) -s \dr \right] \nnb
\\
&&+ m_c  Q_b \left[ - 2 m_b^2 \, \mbox{ln}
\frac{1+\alpha-\beta+\lambda}{1+\alpha-\beta-\lambda}+
\lambda \ga 2 m_b^2 - 2 m_c^2 - p^2 (\alpha -\beta) + s \dr \right] \nnb \\
&&+ m_b  Q_c \left[ 2 m_c^2 \, \mbox{ln}
\frac{1-\alpha+\beta+\lambda}{1-\alpha+\beta-\lambda}+
\lambda \ga 2 m_b^2 - 2 m_c^2 - p^2 (\alpha -\beta) - s \dr \right] \\
&&+ m_c  Q_c \left[ \ga s - p^2 - 2 m_c^2 \dr \mbox{ln}
\frac{1-\alpha+\beta+\lambda}{1-\alpha+\beta-\lambda}-
\lambda \ga 2 m_b^2 - 2 m_c^2 - p^2 (2 + \alpha -\beta) + s \dr \right]
\Bigg\} ~,\nnb
\eea
where $M^2$ is the Borel parameter, $\bar u = 1-u$ and
\bea
s = \frac{\ga m_b + m_c \dr^2 - p^2 \bar u}{u} ~. \nnb
\eea
In obtaining expressions (13) and (14) we have introduced two new
variables
\bea
u &=& \frac{\ga m_b + m_c \dr^2 - p^2}{s-p^2} ~,\nnb \\ 
\Delta &=& \frac{\ga m_b + m_c \dr^2 - p^2}{s_0-p^2}~, \nnb  
\eea
which is equivalent to the
subtraction of higher states. In our analysis we will evaluate the form
factors $g_1$ and $g_2$ at $p^2=m_V^2$.

\section{Numerical analysis}
In our calculation of the form factors $g_1( p^2=m_V^2 )$ and
$g_2( p^2=m_V^2 )$, we use the following set of parameters:
$m_b = 4.7 ~GeV,~m_c=1.4~GeV,~m_{B_c}=6.258~GeV$ \cite{R1,R10}, 
$s_0=50~GeV^2$, $f_\rho = 0.216~GeV$, $f_{K^*} = 0.211~GeV$  
and $f_{B_c}=0.35~GeV$ \cite{R1,R10,R11,R12}. The analysis of
the dependence of $g_1( p^2=m_V^2 )$ and 
$g_2( p^2=m_V^2 )$ on the Borel parameter $M^2$ shows that the
best stability is achieved in the region $15~GeV^2 < M^2 < 20~GeV^2$.
The predictions of the
sum rules on the form factors have errors by at most $10\%$ due to the
uncertainties in $m_b,~s_0$, $f_{B_c}$ and $M^2$ in the above--mentioned
region. Our numerical analysis on the form factors 
$g_1( p^2=m_V^2 )$ and $g_2( p^2=m_V^2 )$ predicts the
following results:
\bea
g_1( p^2=m_\rho^2 )   &=& 0.44 ~GeV,~~~~~~~
g_1( p^2=m_{K^{*+}}^2 )  =  0.44~GeV,\nnb \\
g_2( p^2=m_\rho^2 )   &=& 0.21 ~GeV,~~~~~~~
g_2( p^2=m_{K^{*+}}^2 )  =  0.21 ~GeV.
\eea

The branching ratio of the $B_c \rar V \gamma$ decay is
\bea 
{\cal B} ( B_c \rar V \gamma ) = \frac{G^2 \alpha}{16} 
\vel V_{cb} V_{uq}^* \ver^2 f_V^2 m_V^2 
\ga \frac{m_{B_c}^2-m_V^2}{m_{B_c}} \dr^3 
\left[ \frac{g_1^2 ( m_V^2 )}{m_{B_c}^4} + 
\frac{g_2^2 ( m_V^2 )}{m_{B_c}^4} \right] \tau(B_c)~,
\eea
where for the $B_c$ meson life time we have used 
$\tau(B_c)=0.52 \times 10^{-12}~s$ \cite{R13}, $\vel V_{ud} \ver = 0.97$,
$\vel V_{us} \ver = 0.22$ and 
$\vel V_{cb} \ver = 0.04$ \cite{R14}. With this set of parameters, finally,
we summarize the numerical results of the branching ratios.
\bea
{\cal B} ( B_c \rar \rho^+ \gamma ) &=& 8.3 \times 10^{-8}~, \nnb \\
{\cal B} ( B_c \rar K^{*+} \gamma ) &=& 5.3 \times 10^{-9}~~.
\eea
Few words about the experimental observability of these decays are in order.
In \cite{R15,R16} it is estimated that at LHC, approximately 
$2 \times 10^8~ B_c$
mesons per year will be produced. Using the result of Eq. (18) and this
estimated number of decays, we can easily calculate the number of expected
events for the $B_c \rar V \gamma$ decay at LHC to be
\bea
{\cal N} (B_c \rar \rho^+ \gamma ) &=& {\cal B}(B_c \rar \rho^+ \gamma ) 
\times (2 \times 10^8) = 17~, \nnb \\
{\cal N} (B_c \rar K^{*+} \gamma ) &=& {\cal B}(B_c \rar K^{*+} \gamma )
\times(2 \times 10^8) = 1~. \nnb
\eea

From this estimation it follows that at future LHC collider it is 
possible to detect only $B_c \rar \rho \gamma$ channel.

\end{document}